\documentclass[10pt,letterpaper,compsoc,conference]{iiswc25}

\usepackage{cite}
\usepackage{amsmath,amssymb,amsfonts}
\usepackage{graphicx}
\usepackage[dvipsnames]{xcolor}
\usepackage[final]{microtype}
\usepackage[italic]{mathastext}
\usepackage{libertine}
\usepackage[T1]{fontenc}
\usepackage{textcomp}
\usepackage[varqu,varl]{zi4}
\usepackage[all]{nowidow}
\usepackage[keeplastbox]{flushend}
\usepackage{fancyhdr}
\usepackage[bookmarks=true,breaklinks=true,letterpaper=true,colorlinks,citecolor=blue,linkcolor=blue,urlcolor=blue]{hyperref}

\usepackage{dirtree}

\usepackage{booktabs} 

\usepackage[caption=false,font=normalsize,labelfont=sf,textfont=sf]{subfig}
\usepackage{multirow}

\def\fixme#1{\bgroup \color{red}{[{#1}]}\egroup}

\usepackage[none]{hyphenat} 
\usepackage{tikz}
\DeclareRobustCommand{\circled}[1]{%
  \tikz[baseline=(myanchor.base)] 
    \node[circle,fill=black,inner sep=.5pt] (myanchor) 
    {\color{white}\footnotesize #1};%
}

\usepackage{algpseudocode}
\makeatletter
\renewcommand{\ALG@beginalgorithmic}{\scriptsize}
\makeatother

\usepackage{listings} 
\usepackage{color} 
\definecolor{codegreen}{rgb}{0,0.6,0}
\definecolor{codegray}{rgb}{0.5,0.5,0.5}
\definecolor{codepurple}{rgb}{0.58,0,0.82}
\definecolor{backcolour}{rgb}{1,1,1}

\newcommand{\sixpt}{\fontsize{6.4pt}{7.2pt}\selectfont}

\lstdefinestyle{mystyle}{
    backgroundcolor=\color{backcolour},   
    commentstyle=\itshape\color{codegreen},
    keywordstyle=\color{magenta},
    numberstyle=\tiny\color{codegray},
    stringstyle=\color{codepurple},
    basicstyle=\scriptsize,
    breakatwhitespace=false,         
    breaklines=true,                 
    captionpos=b,                    
    keepspaces=true,                 
    numbers=left,                    
    numbersep=5pt,                  
    showspaces=false,                
    showstringspaces=false,
    showtabs=false,                  
    tabsize=7,
    frame=single
}

\newsavebox{\codebaseline}
\newsavebox{\codeupdated}

\fancypagestyle{firstpage}{
  \fancyhf{}
  
}

\begin{document}


\title{Different Perspectives of Memory System Simulation}





\author{%
  \IEEEauthorblockN{%
    \IEEEauthorrefmark{1}Pouya~Esmaili-Dokht\IEEEauthorrefmark{2}\IEEEauthorrefmark{4}, 
    \IEEEauthorrefmark{1}Arash~Yadegari\IEEEauthorrefmark{4}, 
    Victor~Xirau\IEEEauthorrefmark{4}, 
    Julian~Pavon\IEEEauthorrefmark{4}, 
    Adri\'an~Cristal\IEEEauthorrefmark{2}\IEEEauthorrefmark{4},\\
    Eduard~Ayguad\'e\IEEEauthorrefmark{2}\IEEEauthorrefmark{4}, 
    Petar~Radojkovi\'c\IEEEauthorrefmark{4}%
  }
  \IEEEauthorblockA{%
   Universitat Polit\`ecnica de Catalunya\IEEEauthorrefmark{2}, 
   Barcelona Supercomputing Center\IEEEauthorrefmark{4}, 
  }
  \IEEEauthorblockA{%
    \{pouya.esmaili, arash.yadegari, victor.xirau, julian.pavon, adrian.cristal, eduard.ayguade, petar.radojkovic\}@bsc.es
  }
  \thanks{\IEEEauthorrefmark{1}Both authors contributed equally to this study.}
}

\IEEEoverridecommandlockouts
\maketitle
\thispagestyle{firstpage}

\pagestyle{plain}


\begin{abstract}
Memory simulators are used to estimate application performance on advanced memory systems, yet they may exhibit significant discrepancies compared to real hardware. 
This paper investigates two key questions:\,(1)\,what causes these inaccuracies, and (2)\,how can simulators be properly validated to ensure reliable performance predictions.
We propose a methodology that evaluates memory performance from three complementary perspectives: the memory simulator, the CPU–memory interface, and the application. Our analysis reveals that these perspectives can diverge substantially, with application-level performance often decoupled from internal simulator statistics. We identify the CPU–memory interface as the primary source of these inaccuracies.
To address these problems, we implement a set of corrections and enhancements that improve the fidelity of integrated simulators. We evaluate these changes across multiple widely used simulators, including Ramulator, Ramulator\,2, and DRAMsim3 integrated with ZSim. The results show that correcting interface-related issues is essential to achieve simulation outcomes that closely resemble actual system performance.
\end{abstract}

\section{Introduction}
\label{sec:Introduction}

\looseness -1 For decades, comparing simulated DRAM timings against
manufacturers’ Verilog models has been considered the gold
standard for memory simulator validation\,\cite{steiner:dramsys4, Shangli:dramsim3, kim:ramulator, Haocong:Ramulator2}.
A recent study, however, showed this is not the case, and that even simulators
that do not violate DRAM Verilog timings can show
surprisingly large discrepancies with the actual-system performance\,\cite{esmaili:mess}. This raised two
major questions. First, what is the source of these large simulation 
inaccuracies? Second, how can we extend validation of memory
simulators to ensure they resemble actual hardware
performance? This paper provides some answers to these questions.

\begin{figure}[!t]
 \centering
 \includegraphics[width=\linewidth]{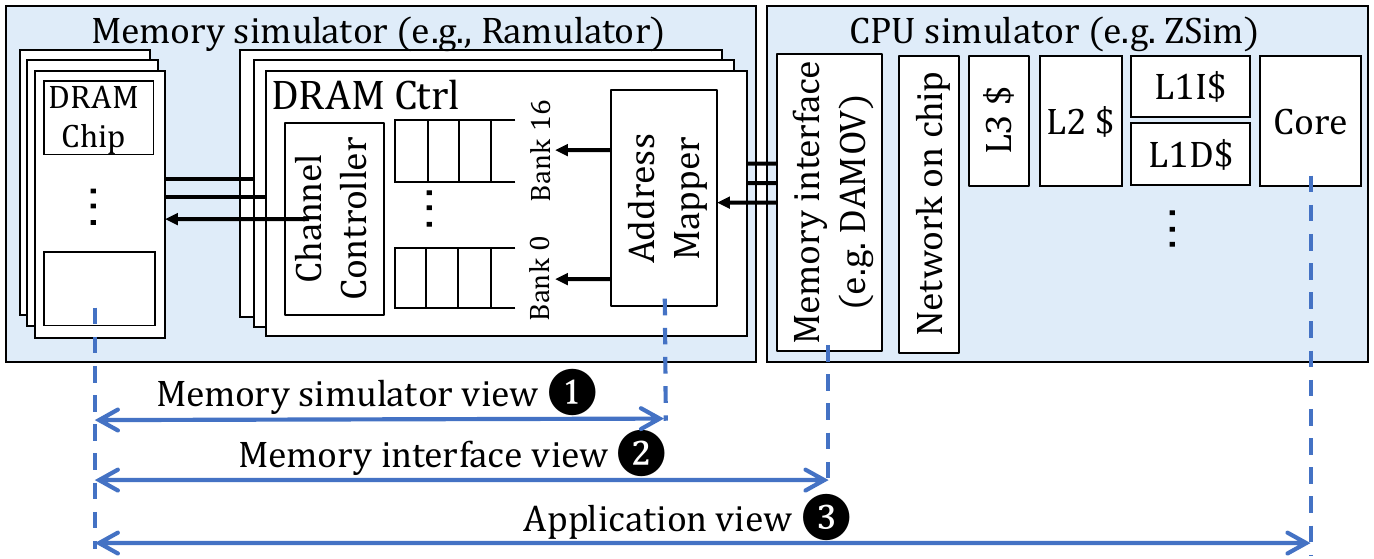}%
 \caption{Memory simulator\,\circled{1} and interface\,\circled{2} views disclose \textbf{internal} simulators' statistics. The application view\,\circled{3} is the \textbf{final memory-simulation outcome} that determines the simulated application performance.}
 \label{fig:3view-methodology}
\end{figure}

In complex simulation platforms that comprise integrated CPU and memory simulators, 
the memory system performance can be evaluated in multiple places of the simulated execution pipeline. 
Fig.\,\ref{fig:3view-methodology} illustrates this with an example of a simulation platform 
that comprises interconnected CPU and memory simulators. 
The figure also shows three memory performance views that are covered in this study.  
%
The \textbf{memory simulator view}\,\circled{1}  
shows how the simulator \textbf{internally} processes memory requests. 
The \textbf{memory interface view}\,\circled{2} monitors the simulation statistics at the 
memory interface of the CPU simulator. 
It shows how the \textbf{CPU simulator perceives} the simulated memory performance. 
Surprisingly, although these two views are nearby in the hardware simulation platform, 
our study shows that they can have a very different memory-performance perception.  
Finally, the \textbf{application view}\,\circled{3} measures 
the\textbf{ application load-to-use latency}. 
While the first two views disclose \textbf{internal} memory-related simulator statistics, 
the application view is the \textbf{simulation outcome} that directly determines the simulated application performance.
For this reason, the application view is the \textbf{ultimate} measure of the memory simulation correctness that should be compared to the actual hardware for validation purposes.

 \looseness -1 Our study shows how detailed evaluation of various memory-performance views 
can help users to \textbf{detect}, \textbf{understand} and \textbf{correct simulation inaccuracies}. 
We also analyze and correct two fundamental problems of integrated simulation infrastructures: 
the CPU- and memory-simulator \textbf{clocking} and the mismatch between the \textbf{immediate-response} and \textbf{cycle-accurate} simulation models (Sec.\,\ref{sec:interface}). 
We also address the issues of 
incorrect \textbf{address mappings}, simplified \textbf{network-on-chip models}, and absence of \textbf{data prefetchers} (Sec.\,\ref{sec:other-source}). 
We deploy and evaluate these corrections with the ZSim CPU simulator connected to Ramulator, Ramulator\,2 and DRAMsim3. 
All simulator enhancements are released as open source~\cite{interface:repo} and are ready to be used by the community, as detailed in Artifact Appendix.

\begin{figure*}[!t]
  \centering
  \includegraphics[width=.4\linewidth]{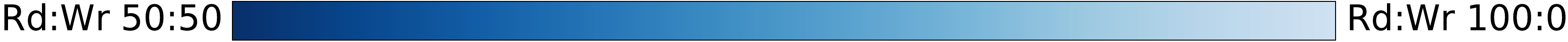}
\vspace{-.3cm}

  \subfloat[Intel\,Skylake\,server\,with 6$\times$DDR4-2666%
           \label{fig:actual-hw}]{%
    \includegraphics[width=.25\linewidth]{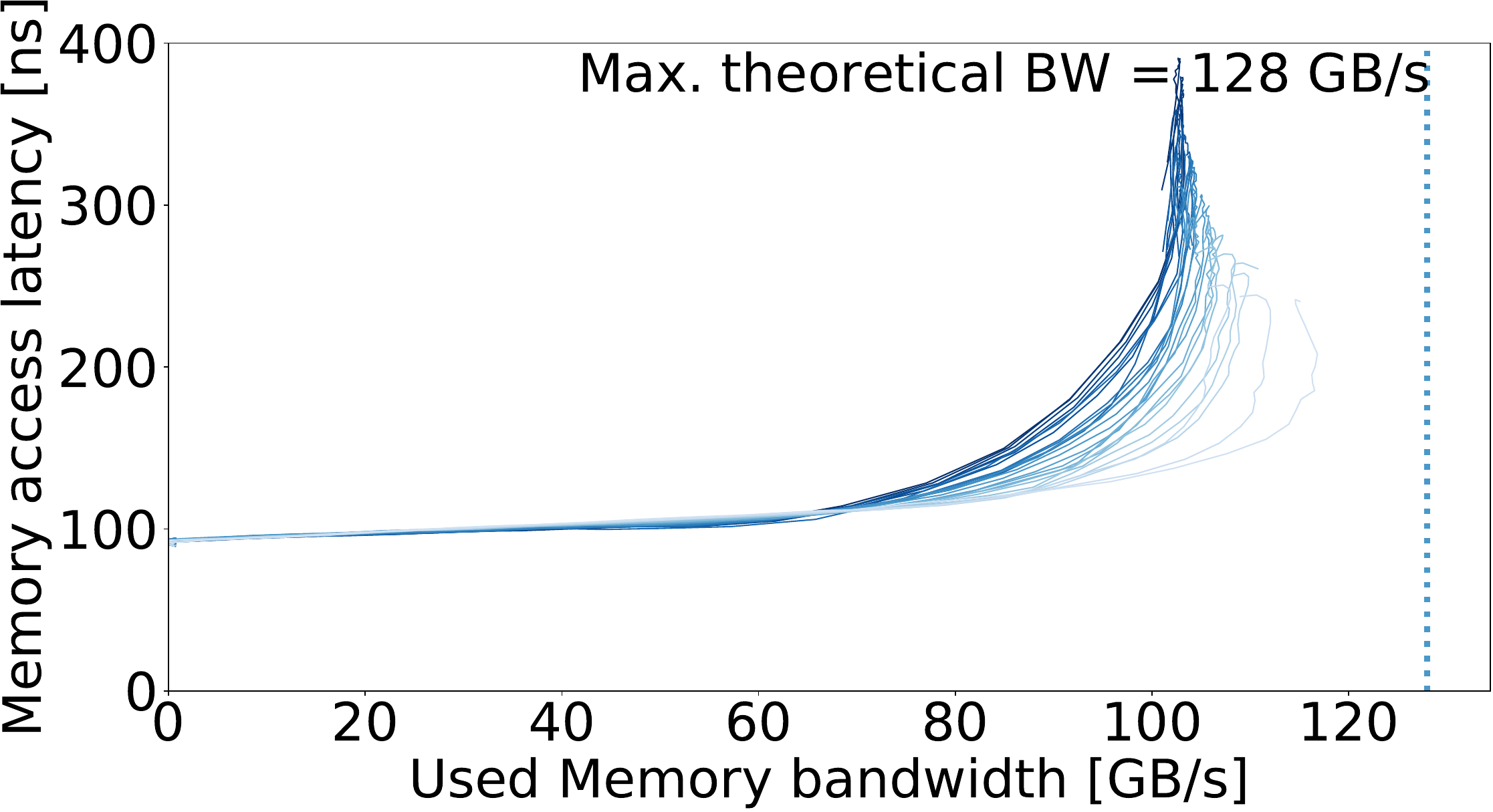}%
  }%
  \hfill
  \subfloat[Memory simulator view%
           \label{fig:baseNoFix_ramulator}]{%
    \includegraphics[width=.25\linewidth]{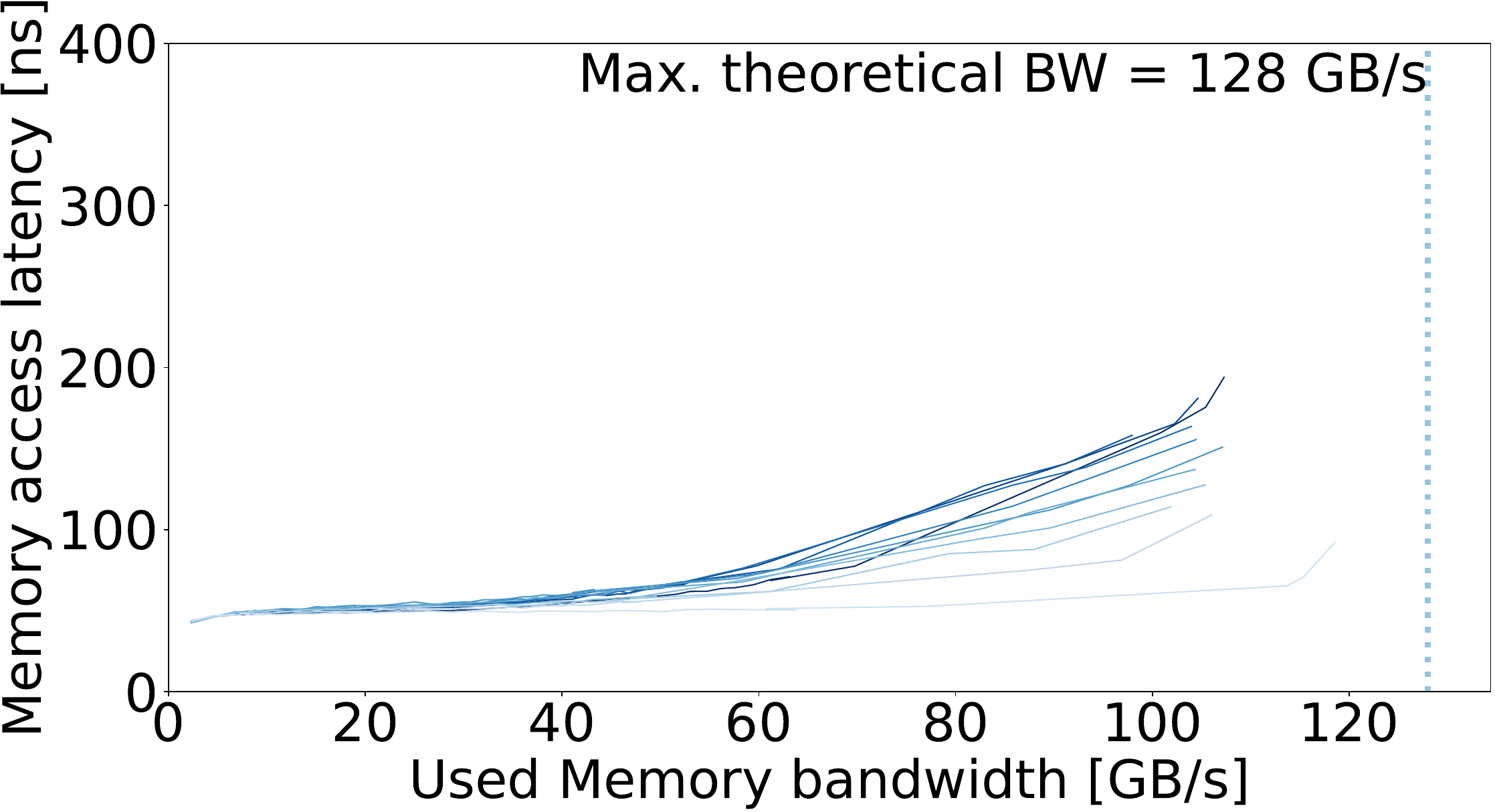}%
  }%
  \subfloat[Memory interface view%
           \label{fig:baseNoFix_interface}]{%
    \includegraphics[width=.25\linewidth]{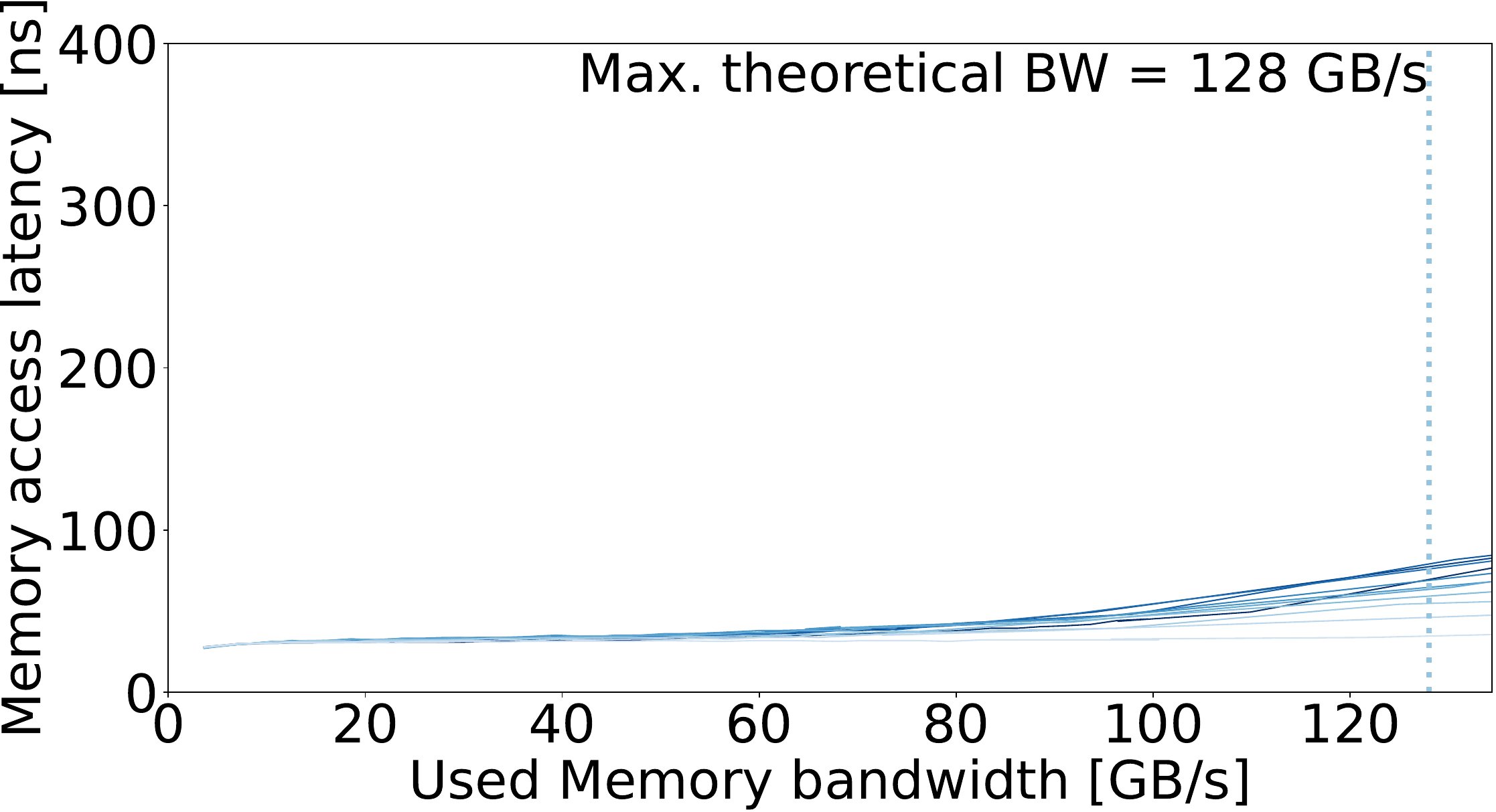}%
  }%
  \hfill
  \subfloat[Application view%
           \label{fig:baseNoFix_core}]{%
    \includegraphics[width=.25\linewidth]{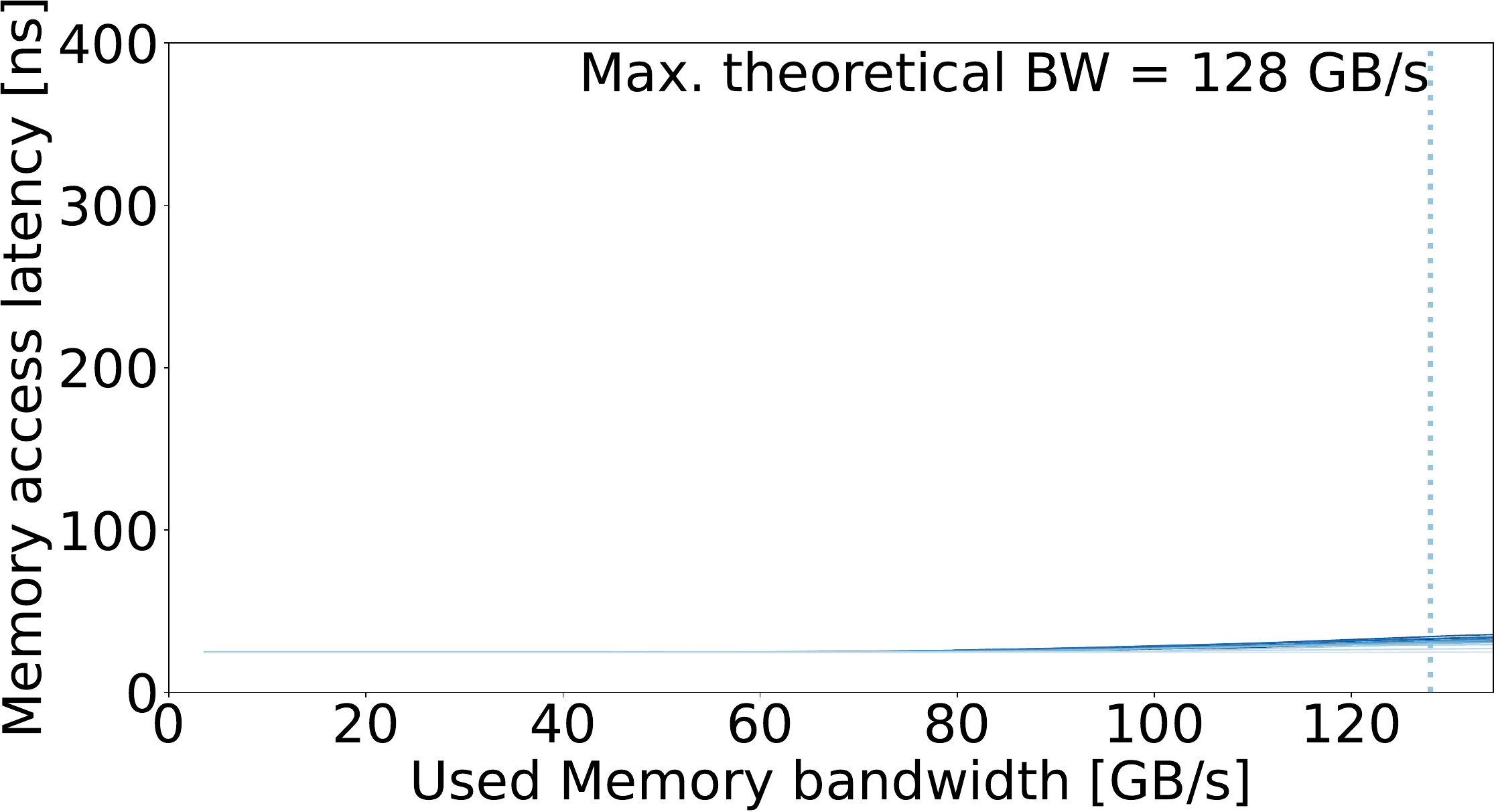}%
  }%

  \caption[Four subfigures]{%
    The application view of the simulated load-to-use latency is 24\,ns across the entire realistic bandwidth range (Fig.\,\ref{fig:baseNoFix_core}). This is completely decoupled from the actual system performance (Fig.\,\ref{fig:actual-hw}), and even more surprisingly, from the  memory simulator itself (Fig.\,\ref{fig:baseNoFix_ramulator}). }  
  \label{fig:original-damov-interface}
\end{figure*}

\section{Experimental environment}
\label{sec:Methodology}

We use a simulation platform that comprises event-based ZSim
CPU simulator~\cite{sanchez:zsim,esmaili:ACM} ,  and cycle-accurate Ramulator~\cite{kim:ramulator}, integrated
with the DAMOV interface~\cite{Oliveira:amov-ZSim-ramulator}, see Fig.\,\ref{fig:3view-methodology}.
In Sec.\,\ref{sec:finalChart} we also integrate and validate ZSim with Ramulator\,2~\cite{Haocong:Ramulator2}, previously not connected to ZSim, and deploy our corrections to ZSim+DRAMsim3~\cite{esmaili:ACM}. 
The main simulation parameters are listed in Table~\ref{table:configuration}. 
They match the hardware features of the actual Intel Skylake server~\cite{bsc:mn4} which is used for the simulator validation.

Different simulation views are compared with the 
Mess benchmark~\cite{esmaili:mess}  
which performs a detailed memory-system profiling. 
The curves with different composition of read and write traffic are plotted with different shades of blue. 
Each curve shows the evolution of the memory latency ($y$-axis) for different used bandwidth ($x$-axis). 
Fig.\,\ref{fig:actual-hw} shows the Mess characterization of the actual Intel Skylake server used in this study.

\section{Interface: Where CPU and memory meet}
\label{sec:interface}

Interfaces between the CPU and memory simulators determine
how different components interact in a full-system simulation,
playing a critical role in overall simulation accuracy. CPU
and memory simulators are often developed independently by
different research groups, and they embody different design
trade-offs and abstraction levels, such as the event-based CPU
and cycle-accurate memory simulation. Their integration is,
therefore, inherently complex and requires a good understanding
of the simulation infrastructure. Given the importance and
complexity of the simulator interfaces, we argue that more attention
should be paid to their validation. Our study makes important
steps in this direction and demonstrates how evaluation of various
memory-performance views presented in Fig.\,\ref{fig:3view-methodology} can be used to
detect, understand and correct sources of major simulation
inaccuracies.


\subsection{Detected: Interface errors}
\label{sec:detect}
Fig.\,\ref{fig:original-damov-interface} shows the memory-system performance of the actual Intel Skylake server
and different performance views of ZSim~\cite{sanchez:zsim, esmaili:ACM} 
integrated to Ramulator~\cite{kim:ramulator} via the DAMOV interface~\cite{Oliveira:amov-ZSim-ramulator}. 
%
%
%
The \textbf{memory simulator} view (Fig.\,\ref{fig:baseNoFix_ramulator}), 
with some variations, follows a general trend of the actual-system performance.  
The simulated bandwidth utilization is similar to the actual system, saturating the memory bandwidth between 100\,GB/s
and 120\,GB/s. The unloaded memory latency is 43\,ns, which is below the 89\,ns measured in the actual system. 
However, this is partially because 
the actual system measurements show the whole load-to-use latency, while the memory simulator view 
reports round-trip latency measured from the memory controller.  
Another difference is that, in the saturated memory system (100--120\,GB/s),  
the maximum simulated latency ranges between 90 and 195\,ns which is more than 2$\times$ below the actual 240--390\,ns measurements.  

The \textbf{memory interface} view (Fig.\,\ref{fig:baseNoFix_interface}) shows surprisingly different results. 
First, the memory latency starts at 27\,ns and is notably \textbf{below} the memory simulator view in the whole bandwidth range. Second, the simulated memory bandwidth \textbf{exceeds the maximum theoretical} by 40\%.  

\begin{table}[!t]
\centering
\caption{Parameters of the  simulation platform used in the study}
\label{table:configuration}
\begin{tabular*}{\linewidth}{@{\extracolsep{\fill}}ll}
\toprule
\textbf{ZSim: CPU parameters}   & \textbf{Ramulator: Memory  parameters}    \\
\midrule
Skylake microarchitecture       &	6$\times$ DDR4-2666~channels		         \\
24\,cores @ 2.1\,GHz          	&	1$\times$ 32\,GB DIMM per channel	  	 \\
~~~Private 32\,KB L1-D cache    &	2 ranks per DIMM 	                     \\
~~~Private 32\,KB L1-I cache    &	8 devices per rank                       \\
~~~Private 1\,MB L2 cache    	&	16 banks per device                      \\
Shared 33\,MB LLC cache     	&   128\,MB per device bank	                 \\

\bottomrule
\end{tabular*}
\end{table}

Finally, the \textbf{application view} (Fig.\,\ref{fig:baseNoFix_core}) shows an even higher discrepancy from the actual 
memory system performance. The simulated load-to-use latency is 24\,ns across the entire realistic bandwidth range, 
and the maximum simulated bandwidth significantly exceeds the theoretical maximum.

Overall, Fig.\,\ref{fig:original-damov-interface} leads to a surprising finding:  
The application load-to-use latency is significantly decoupled from the performance of the memory simulator itself. 
We also see serious discrepancies between all three different memory-performance views, 
even between the memory simulator and memory interface, 
which are very close in the simulated architecture, \circled{1} and \circled{2} in Fig.\,\ref{fig:3view-methodology}.

\subsection{Corrected: CPU and memory simulator clocking}
\label{sec:correct}
Different memory-performance views are crucial to detect
the exact source of simulation inaccuracies. For example,
significant discrepancies between the memory simulator (Fig.\,\ref{fig:baseNoFix_ramulator}) and
interface view (Fig.\,\ref{fig:baseNoFix_interface}) indicate that the CPU–memory interface is
improperly handling memory requests.

\begin{figure}[!t]
  \centering
  \includegraphics[width=.8\linewidth]{graphics/horizontal-legend}
\vspace{-.3cm}

  \subfloat[Memory interface view%
           \label{fig:baseline-slightly-modify-interface}]{%
    \includegraphics[width=.5\linewidth]{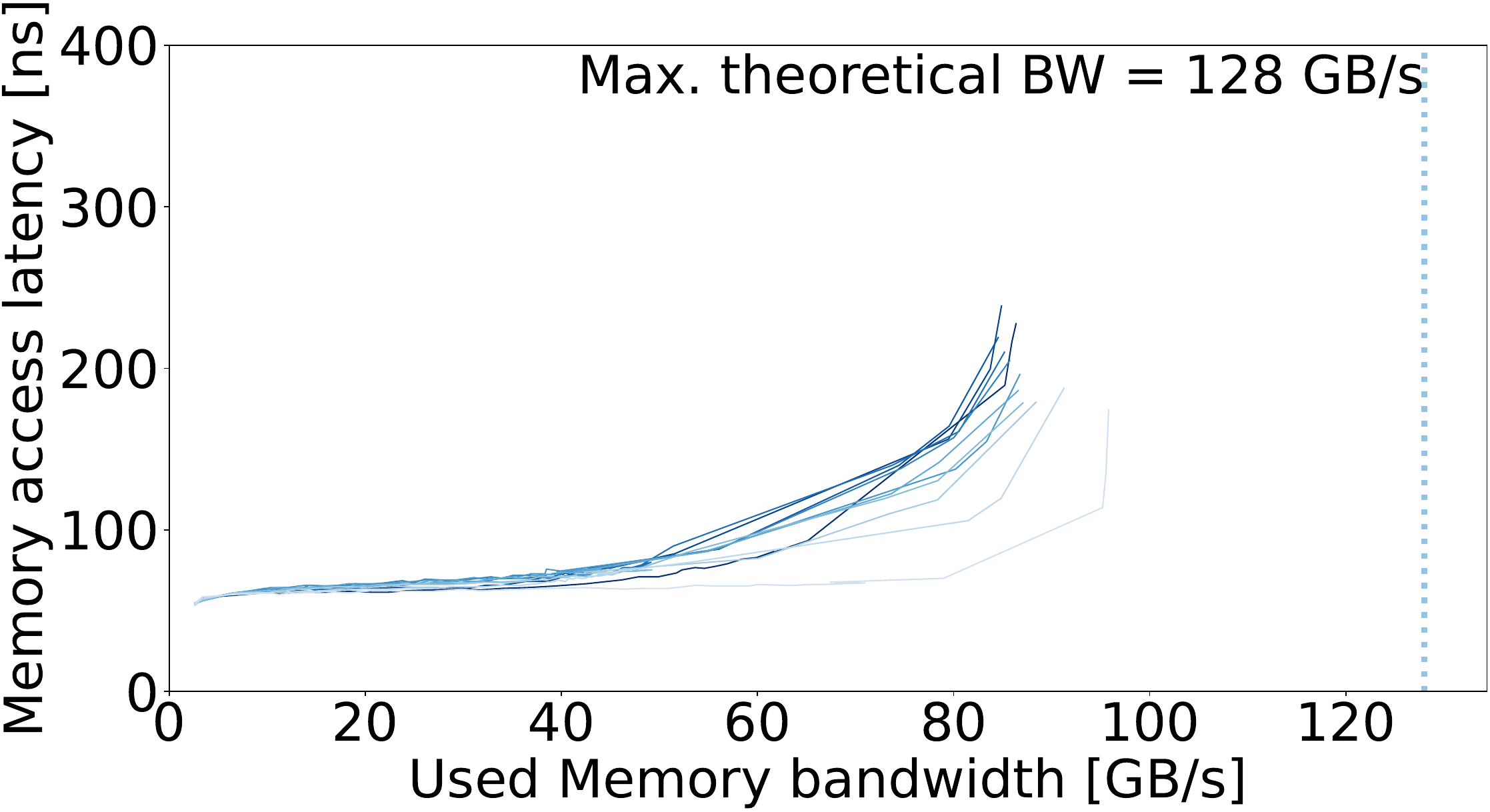}%
  }%
  \subfloat[Application view%
           \label{fig:baseline-slightly-modify-core}]{%
    \includegraphics[width=.5\linewidth]{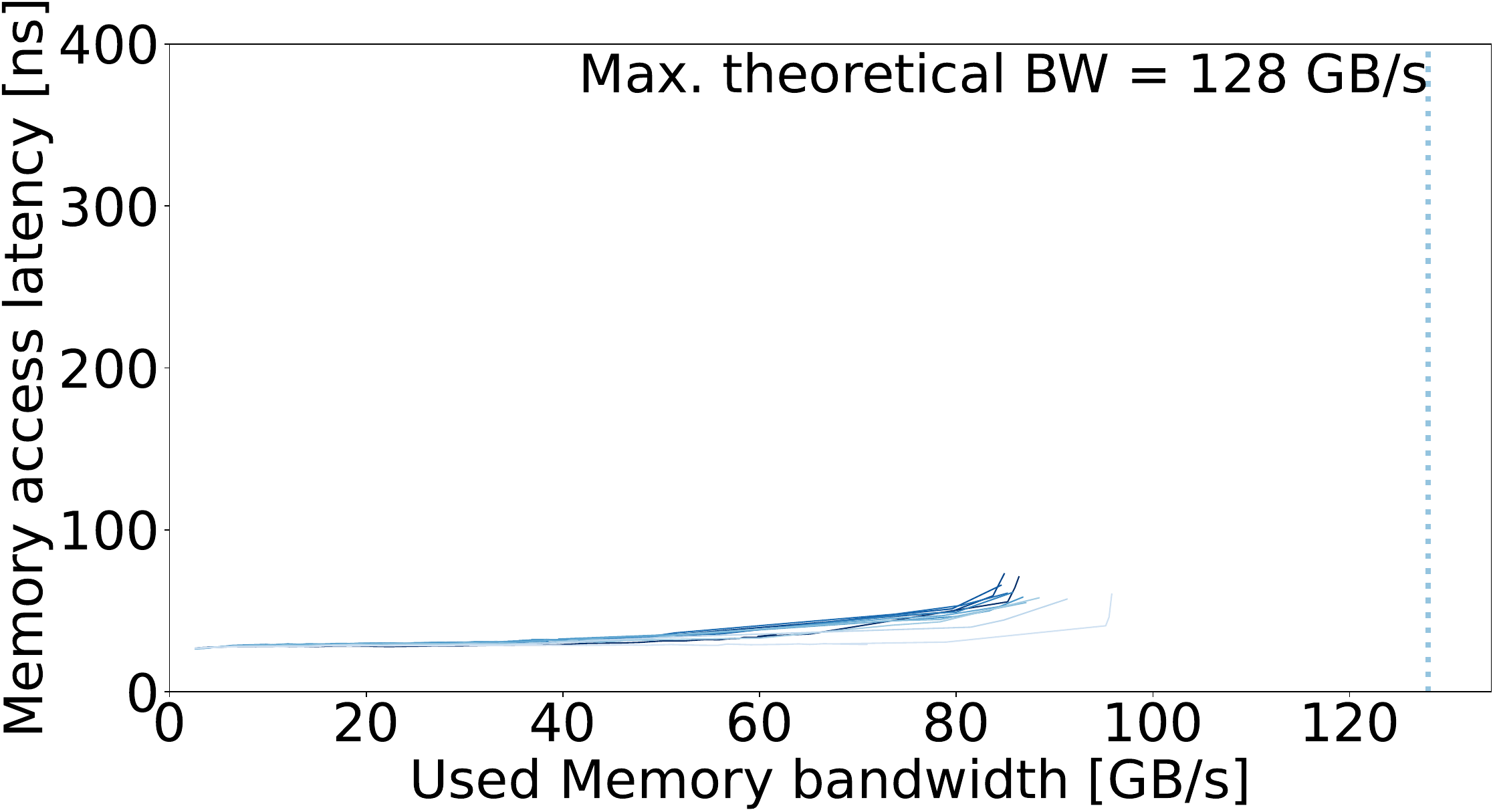}%
  }%
  
  
  \caption[Four subfigures]{%
  Clock-scaling in the memory simulator fixed the unrealistically-high memory bandwidth
  in the memory interface and the application view.} 

  \label{fig:baseline-slightly-modify}
\end{figure}

An inspection of the interface implementation revealed that one
of the inaccuracies originated from the DAMOV interface block responsible for time-domain
consistency between CPU and memory simulators. 
As memory operates at a lower frequency (1.333\,GHz) than the CPU (2.1\,GHz), 
its simulated cycle time (tick) should be proportionally 1.575$\times$ higher. 
In the current simulator release~\cite{safari:damov-template},   
the block responsible for cross-simulator clocking was disabled, 
causing the CPU simulator to perceive the memory system running on 1.575$\times$ higher frequency. 
This caused unrealistically-high memory bandwidth on the CPU side of the memory interface (Fig.\,\ref{fig:baseNoFix_interface}) 
that was further propagated to the application view (Fig.\,\ref{fig:baseNoFix_core}). 
Enabling the clock-scaling code 
corrected the excessive memory system bandwidth (Fig.\,\ref{fig:baseline-slightly-modify}). 

This, however, did not eliminate all memory bandwidth discrepancies, 
and the interface view (Fig.\,\ref{fig:baseline-slightly-modify-interface}) still shows around 25\% lower bandwidth w.r.t. the memory simulator (Fig.\,\ref{fig:baseNoFix_ramulator}).   
We discovered that the second cross-simulator clocking 
issue is caused by the clocking mechanism implemented in the DAMOV interface.   
The interface decides whether to call the 
memory simulator 
depending on the calculated ratio between the CPU and memory frequency, \texttt{freqRatio} in Code Listing~\ref{fig:interface-freq-model}(a).   
The inherent problem of this approach is that \texttt{freqRatio} has to be an integer number, which DAMOV calculates by rounding-up the actual CPU-to-memory frequency ratio.  
%
Because of this rounding error, 
the memory frequency is reduced 2$\times$ instead of 1.575$\times$, 
leading to a proportional memory-bandwidth discrepancy.

\begin{figure}[!t]
  \centering
\includegraphics[width=.8\linewidth]{graphics/horizontal-legend}
\vspace{-.3cm}

  \subfloat[Memory interface view%
           \label{fig:correct-tick-interface}]{%
    \includegraphics[width=.5\linewidth]{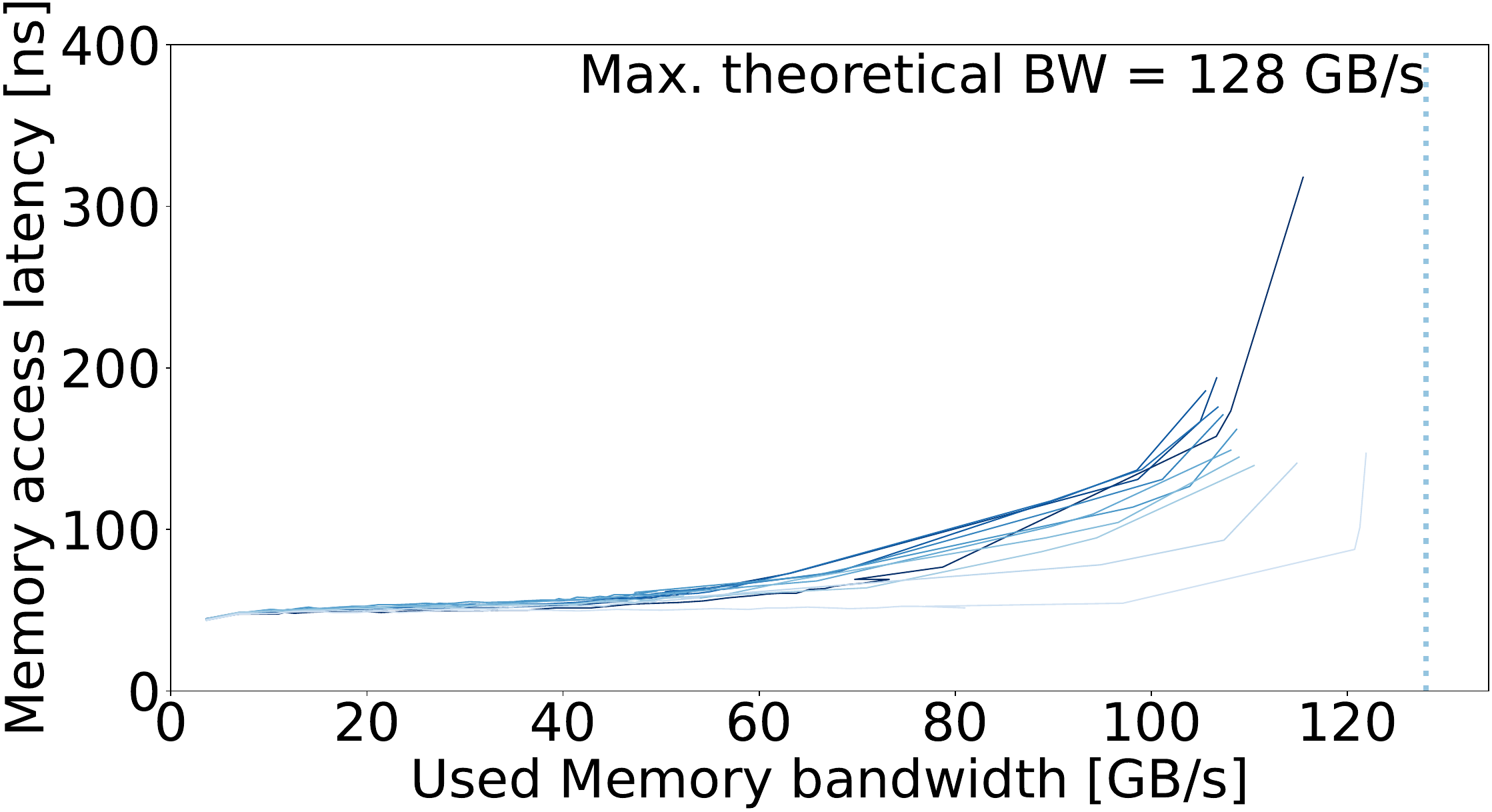}%
  }%
  \subfloat[Application view%
           \label{fig:correct-tick-core}]{%
    \includegraphics[width=.5\linewidth]{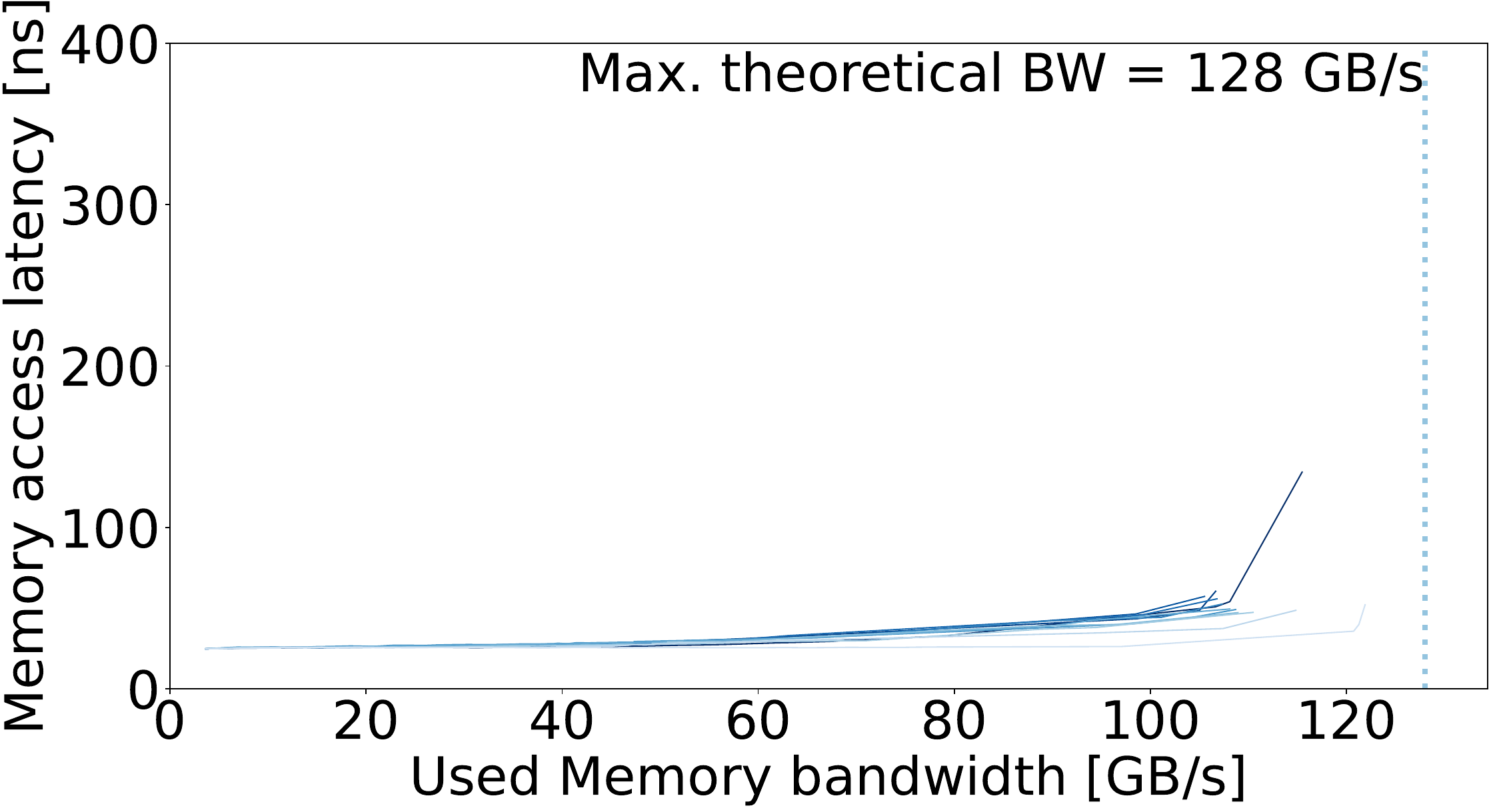}%
  }%

  \caption[Four subfigures]{%
    The updated interface removes the DAMOV \texttt{freqRatio} rounding error. The maximum simulated bandwidth now matches the actual one (Fig.\,\ref{fig:actual-hw}).}

  \label{fig:baseline-clockfixed}
\end{figure}

To address this, we develop a cross-simulator clocking approach, 
presented in Code Listing~\ref{fig:interface-freq-model}(b),
that adjusts the CPU and memory simulated time. 
The approach, first increases the simulated CPU cycles and time (lines~1 and 2).  
Then, while the simulated memory time lags behind the CPU (line~3),  
the interface calls the memory simulator 
tick function, and increases the DRAM time and cycles (lines 4--6).  
The simulated CPU and memory time are increased in time-intervals that correspond to the 
length of their clock cycles in picoseconds: 476\,ps ($\frac{1}{2.1GHz}$) for CPU and 750\,ps ($\frac{1}{1.333GHz}$) for DRAM.
The updated interface removes the DAMOV rounding error and the
maximum simulated memory bandwidth in  
Fig.\,\ref{fig:correct-tick-interface} and 
Fig.\,\ref{fig:correct-tick-core} 
now matches the actual one.

\newcounter{tempFigure} 
\setcounter{tempFigure}{\value{figure}} 
\setcounter{figure}{0} 

\begin{lrbox}{\codebaseline}
\begin{minipage}{0.25\textwidth}
\begin{lstlisting}[basicstyle=\ttfamily\sixpt, frame=single,framerule=0.8pt, rulecolor={\color[RGB]{220, 20, 60}},numbers=left, numberstyle=\tiny, language=C++]
freqRatio = ceil(cpuFreq/memFreq);
if((tickCounter % freqRatio) == 0)
    wrapper->tick();
tickCounter++;         
\end{lstlisting}
\end{minipage}
\end{lrbox}

\begin{lrbox}{\codeupdated}
\begin{minipage}{0.20\textwidth}
\begin{lstlisting}[basicstyle=\ttfamily\sixpt, frame=single,framerule=0.8pt, rulecolor={\color[rgb]{0.0, 0.4, 0.0}}, numbers=left, numberstyle=\tiny, language=C++]
curCycle++;           
cpuPs += cpuPsPerClk;           
while (cpuPs > dramPs) {      
    wrapper->tick();       
    dramPs += dramPsPerClk;
    dramCycle++;
}
\end{lstlisting}
\end{minipage}
\end{lrbox}

\begin{figure}[!t]

    \renewcommand{\figurename}{Listing} 
    \centering
    \subfloat[DAMOV clocking mechanism: \texttt{freqRatio} has to be an integer number, which DAMOV calculates by
rounding-up the actual CPU-to-memory frequency ratio.     
    \label{lst_baseline}]{\raisebox{0.38cm}{\usebox{\codebaseline}}}
    \hfill
    \subfloat[Updated interface \label{lst_updated}]{\usebox{\codeupdated}}
    \caption{The updated interface avoids DAMOV \texttt{freqRatio} rounding error}
    \label{fig:interface-freq-model}
\end{figure}

\setcounter{figure}{\value{tempFigure}}

\subsection{Detected: Memory model mismatch} 
\label{CPU–Memory-Model-Mismatch}
Next, we 
address the discrepancy between the actual memory latency 
(Fig.\,\ref{fig:actual-hw})  
and its application view (Fig.\,\ref{fig:baseline-slightly-modify-core}). 
This large error is caused by a fundamental mismatch between 
\emph{immediate-response} memory model in the first phase of the ZSim simulation, and Ramulator \emph{cycle-accurate} memory simulation~\cite{Shang:misscycle,Eyerman:IRMM}.   

ZSim simulation is done in windows of 1000 cycles and, 
to increase the simulation speed, 
each window is simulated in two phases.   
The first phase simulates non-memory instructions as fast as possible and generates a trace of memory
instructions.  
In the second phase, the memory instructions timings are adjusted based on cycle-accurate Ramulator latency estimates.  
 
The first simulation phase (Bound phase) uses an immediate-response fixed-latency memory model. 
In the DAMOV interface used in the study, this latency is set to a single CPU cycle~\cite{safari:damov-template}.   
This means that, even if an instruction has a dependency on a memory access, it will be issued in the next cycle.      
The second ZSim phase corrects the issue times of these dependent instructions. 
The problem is, however, that the issue times cannot be adjusted for the memory requests that are already passed to the memory simulator. 
The simulator, therefore, practically overlaps these dependent memory accesses,  
leading to an unrealistically-low simulated memory latency.  

This analysis explains the results in Fig.\,\ref{fig:original-damov-interface} and~\ref{fig:baseline-slightly-modify}, 
and reveals an important finding. 
The internal simulator statistics in the memory simulator can \textbf{differ significantly} from the application view of memory latency.     
For this reason, it is \textbf{fundamental} that the simulator developers and users validate their platforms
with performance \textbf{reported by benchmarks} such as the pointer-chase, STREAM~\cite{mccalpin:streamLink} or Mess~\cite{esmaili:mess}. 
Our study shows that internal simulator statistics are not only insufficient, but could be misleading. 

\subsection{Corrected: Memory model mismatch}
\begin{figure}[!t]
  \centering
\includegraphics[width=.8\linewidth]{graphics/horizontal-legend}
\vspace{-.3cm}

  \subfloat[Memory simulator view%
           \label{fig:correct-interface-virgin-ram}]{%
    \includegraphics[width=.5\linewidth]{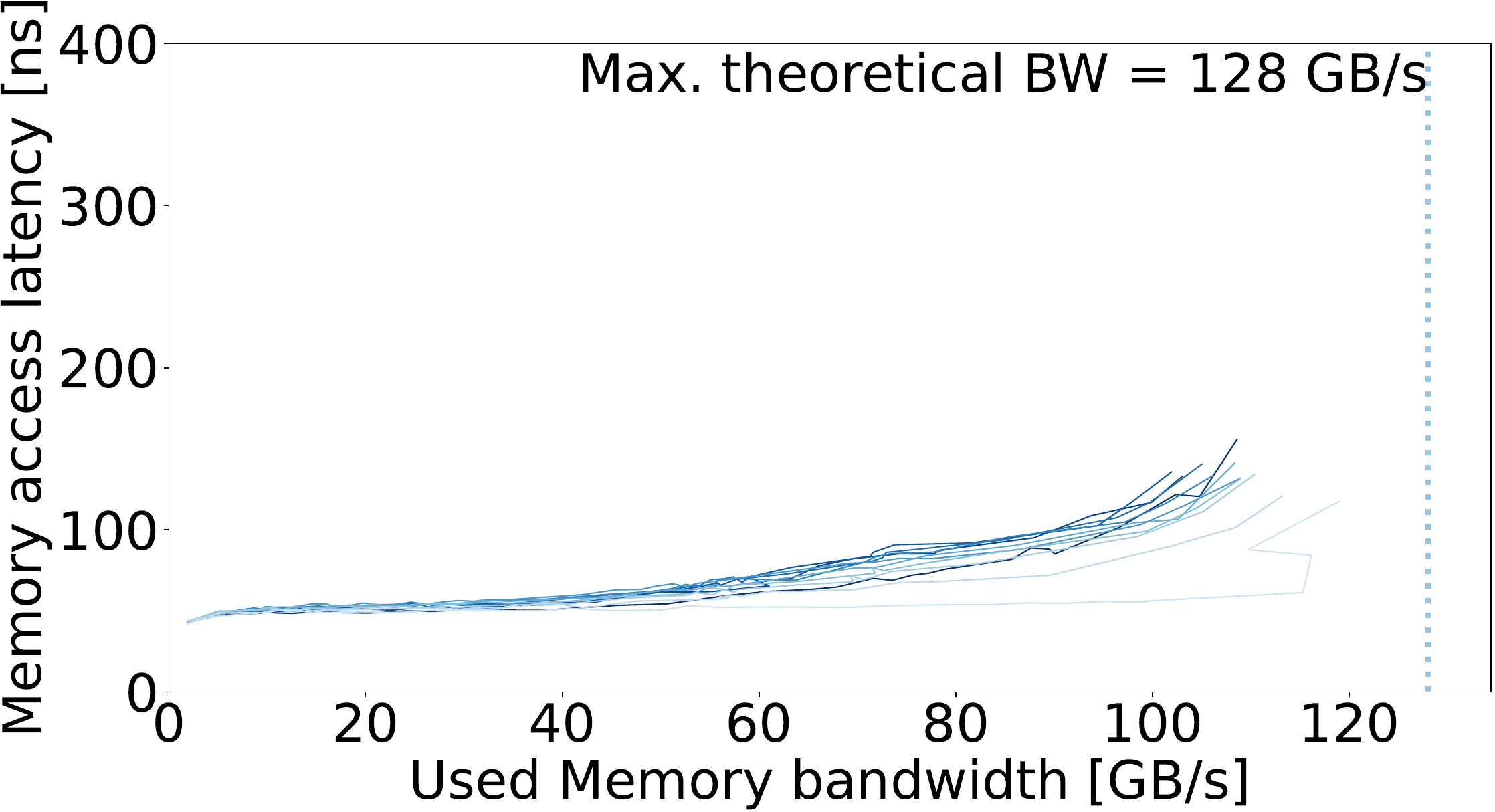}%
  }%
  \subfloat[Application view%
           \label{fig:correct-interface-virgin-core}]{%
    \includegraphics[width=.5\linewidth]{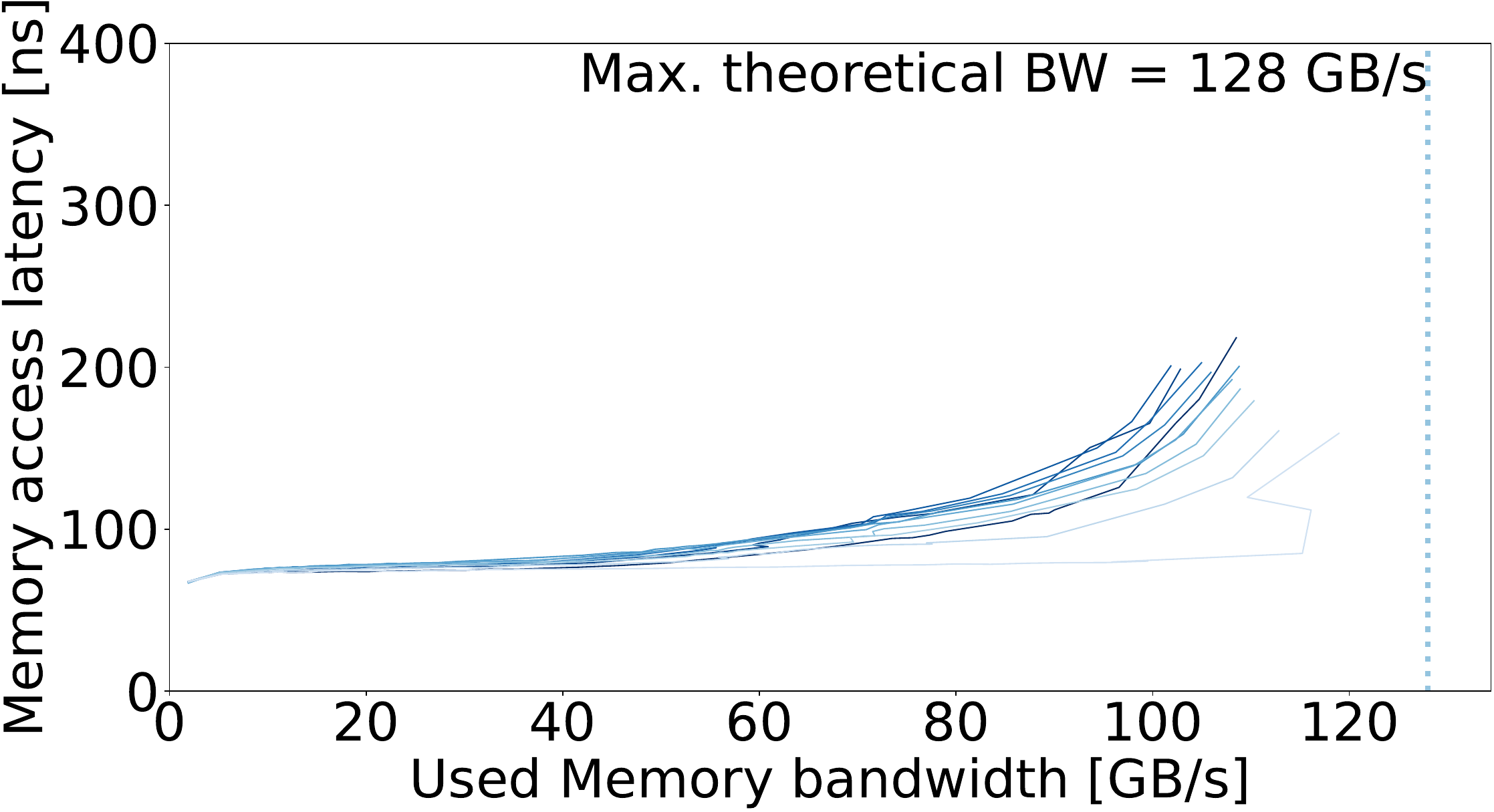}%
  }%

  \caption[Four subfigures]{%
    After the corrections in the simulator interfaces, the application view starts to resemble the outcome of the memory simulator.} 
  \label{fig:correct-interface-virgin}
\end{figure}

The smaller the gap between the simulated latency in the immediate-response and cycle-accurate simulation phases, the smaller the simulation error due to the memory model mismatch.  
%
One way to reduce this gap is to use a proportional--integral controller mechanism~\cite{Franklin:PIDController} from the classical control theory to dynamically update the immediate-response memory latency in the CPU--memory interface. 
At the beginning of each simulation window, we calculate this latency as the weighted sum of two factors: 
the previous immediate-response estimate (95\% weight) and the average memory latency from the previous 
cycle-accurate memory simulation phase (5\% weight).
This keeps a low gap between the memory latency in the two ZSim phases and mitigates the simulation error:  
the unloaded latency improves from 25\,ns (Fig.\,\ref{fig:correct-tick-core}) to 67\,ns (Fig.\,\ref{fig:correct-interface-virgin-core}). 
More importantly, the application view starts to resemble the outcome of the memory simulator (Fig.\,\ref{fig:correct-interface-virgin-ram}) 
and the actual memory-system performance (Fig.\,\ref{fig:actual-hw}).   
%

\subsection{Broader applicability}

The detected simulation inaccuracies and proposed corrections 
are not specific to ZSim and Ramulator.    
Most of the commonly-used CPU and GPU simulators use asynchronous memory model\,\cite{Eyerman:IRMM}, 
while cycle-accurate memory simulators are de facto standard. 
In integrated simulation platforms, the simulated processor and memory clock have to be synchronized.  
Our work demonstrated that this synchronization is not trivial and prone to errors that can devastate the simulation accuracy. 
Also, the analyzed mismatch between the ZSim and Ramulator memory models  
is just an example of the fundamental mismatch between the immediate-response and cycle-accurate memory simulation. This issue concerns numerous CPU and practically all detailed memory simulators~\cite{Shang:misscycle, Eyerman:IRMM}. 
In Sec.\,\ref{sec:finalChart}, 
we will validate our final ZSim memory interface with 
Ramulator\,2~\cite{Haocong:Ramulator2} and DRAMsim3~\cite{Shangli:dramsim3}.

\begin{figure}[!t]
  \centering
	\includegraphics[width=.8\linewidth]{graphics/horizontal-legend}
\vspace{-.3cm}

  \subfloat[Correct address mapping%
           \label{fig:final-am-core}]{%
    \includegraphics[width=.5\linewidth]{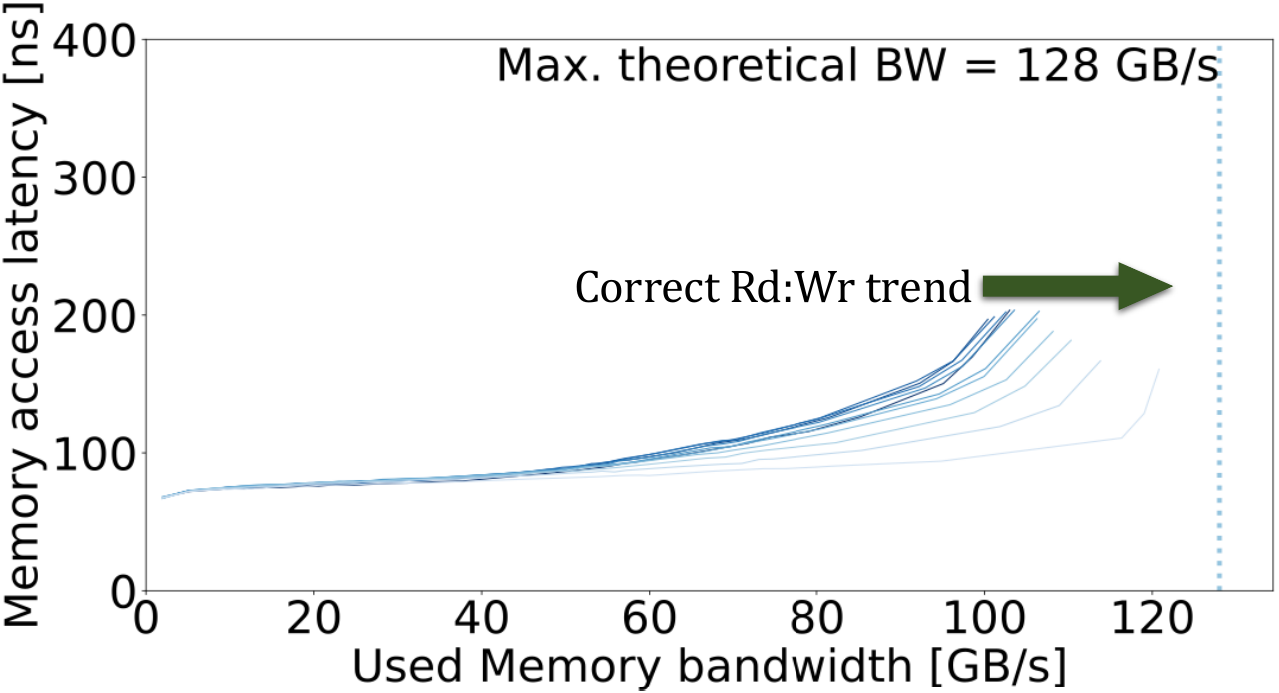}%
  }%
  \subfloat[Realistic network-on-chip%
           \label{fig:final-addnoc-core}]{%
    \includegraphics[width=.5\linewidth]{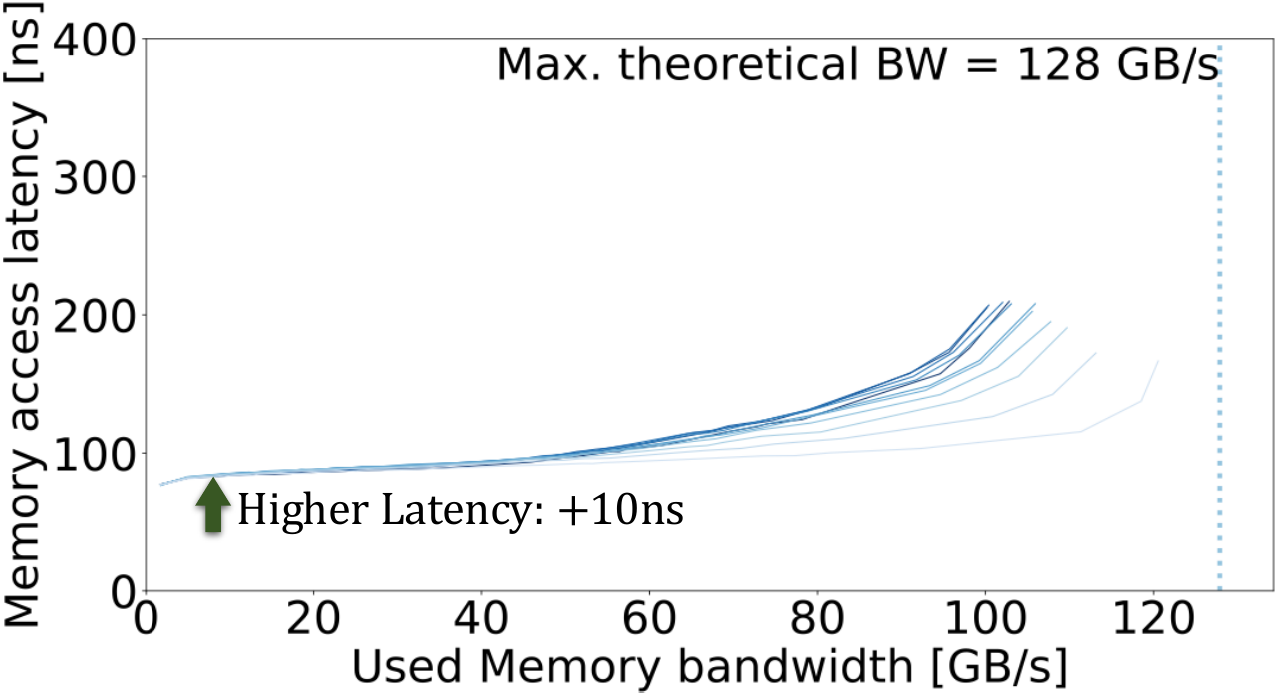}%
  }%

  \subfloat[Adding data prefetcher%
           \label{fig:prefetcher}]{%
    \includegraphics[width=.5\linewidth]{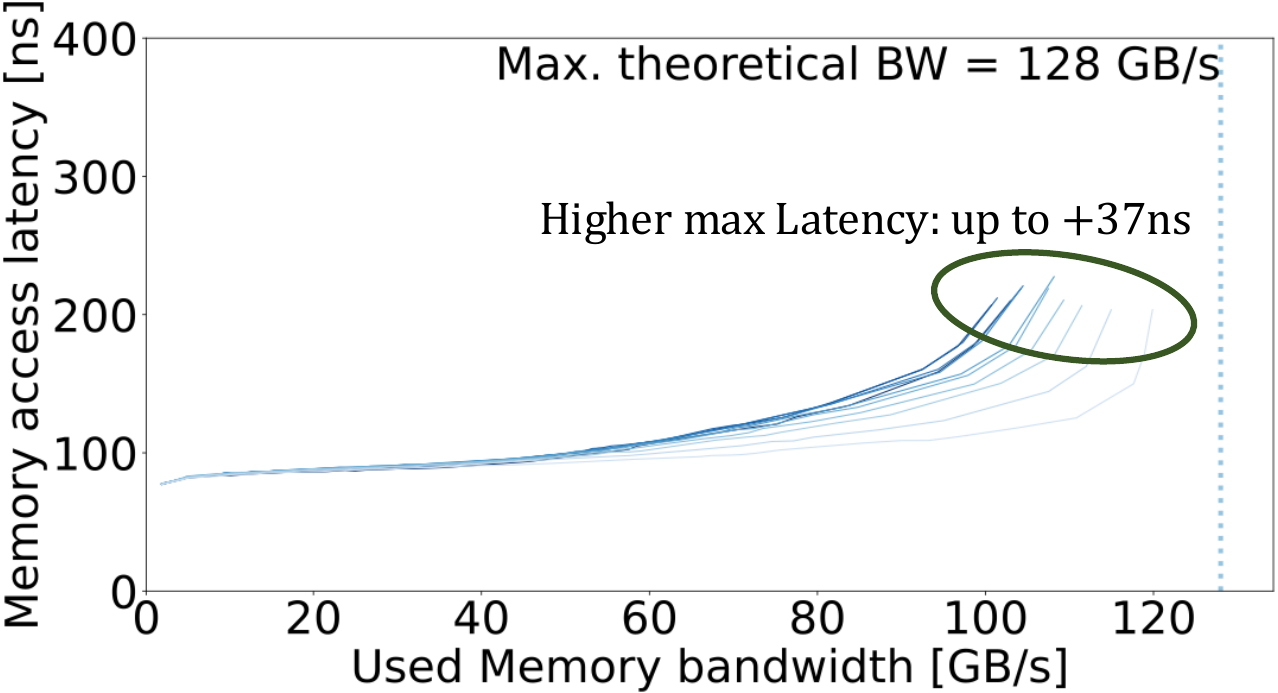}%
  }%
  
  \caption[Four subfigures]{%
    Close-to-hardware accuracy of memory simulation also requires correct address mappings, detailed network-on-chip models and data prefetchers. }

  \label{fig:ZSim-ramulator-final-modifications}
\end{figure}

\section{Further enhancements}
\label{sec:other-source}

In addition to 
CPU--memory simulator interface, 
we correct other frequent sources of simulation inaccuracies and
evaluate their impact on the memory-simulation accuracy.  

The first correction is related to the performance impact of read and write memory traffic. 
In the actual system, the lowest latency and the highest achieved bandwidth are obtained for 100\%-read traffic. 
Memory writes reduce the memory performance and reach the saturation point sooner. 
The higher the percentage of writes, the higher the performance impact,  
and the actual system follows a clear gradient from the lightest (100\%-read) to darker memory curves, 
as seen in Fig.\,\ref{fig:actual-hw}. 
This trend is not reflected in the memory simulator and application views in Fig.\,\ref{fig:correct-interface-virgin}. 
We discovered that the source of this discrepancy is the memory simulator's simplified \textbf{mapping 
between the physical address and the actual data location}: memory channel, DIMM,  
rank, bank, row and column. 
%
After we deployed a complex mapping reverse-engineered from actual systems~\cite{Wang:addressMapping},   
memory curves with different read and write traffic, shown in Fig.\,\ref{fig:final-am-core},  closely follow the actual trends. 

The load-to-use latency is also impacted by CPU \textbf{network-on-chip~(NOC)}.   
Our baseline simulation platform uses a simple NOC modeled with a fixed-delay which is included in the overall LLC latency~\cite{esmaili:ACM}.
We extend the ZSim with 
an more realistic NOC module~\cite{Oliveira:amov-ZSim-ramulator} and then configure it 
so its architecture~\cite{dai:reverseNOC}, latencies~\cite{sodani:noclatency}
and the position of cores in each network tile~\cite{mccalpin:mapping} 
closely match the actual Intel Skylake CPU. 
This adds 10\,ns to the memory latency 
in the whole bandwidth range, see Fig.\,\ref{fig:final-addnoc-core}.


Finally, \textbf{data prefetchers} have an important impact on the memory traffic intensity.  
We implement stride prefetchers in ZSim cores. 
This increases the overall memory traffic resulting in      
up to 37\,ns higher saturated memory latency (Fig.\,\ref{fig:prefetcher}).

\section{Further deployment and future explorations}
\label{sec:finalChart}

\begin{figure}[!t]
\centering
\includegraphics[width=.8\linewidth]{graphics/horizontal-legend}
\vspace{-.3cm}


  \subfloat[ZSim+Ramulator\,2%
           \label{fig:gem5-bw-lat-result-characterization-internal}]{%
    \includegraphics[width=.5\linewidth]{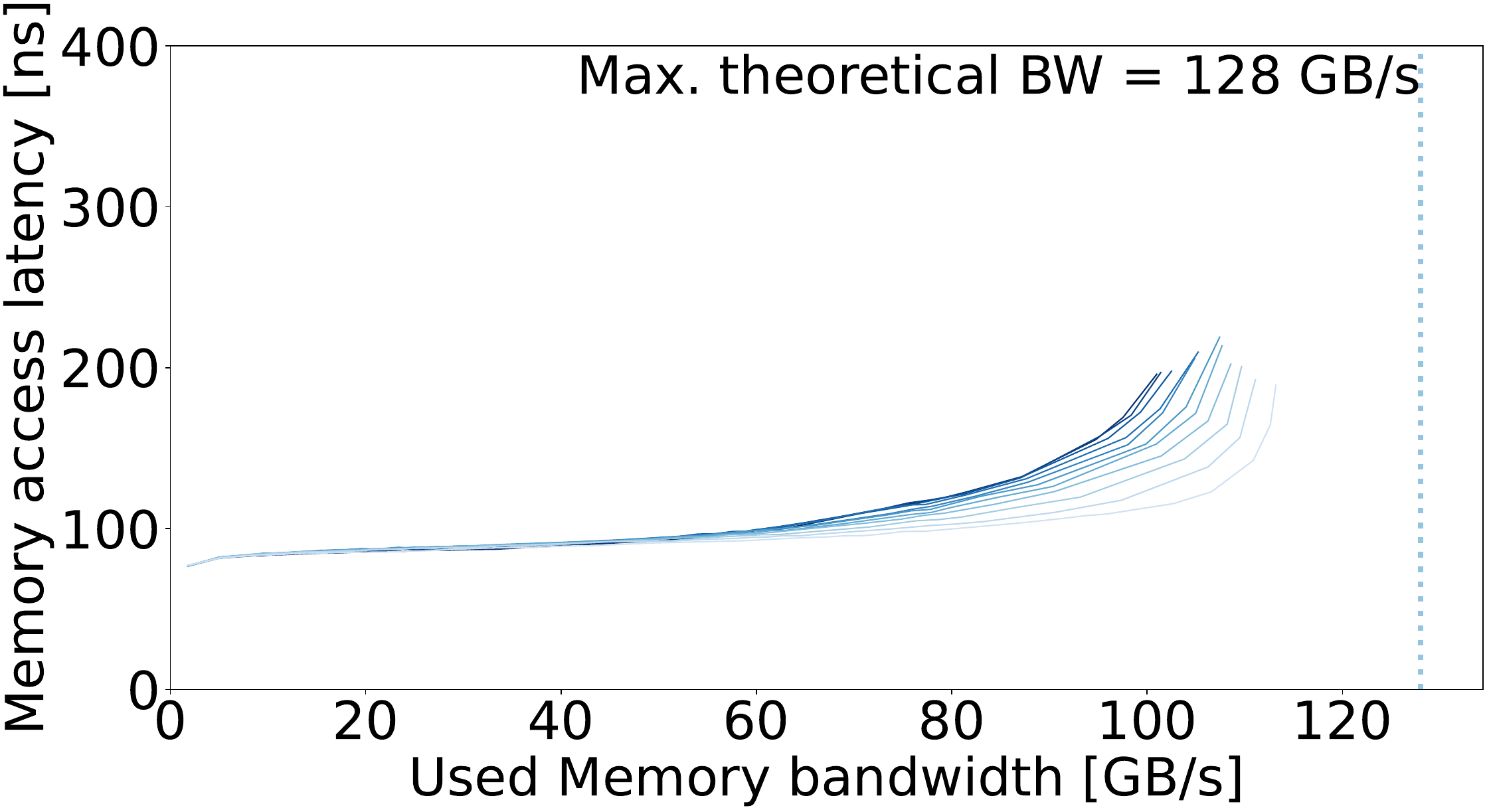}%
  }%
  \subfloat[ZSim+DRAMsim3%
           \label{fig:gem5-bw-lat-result-characterization-ramulator2}]{%
    \includegraphics[width=.5\linewidth]{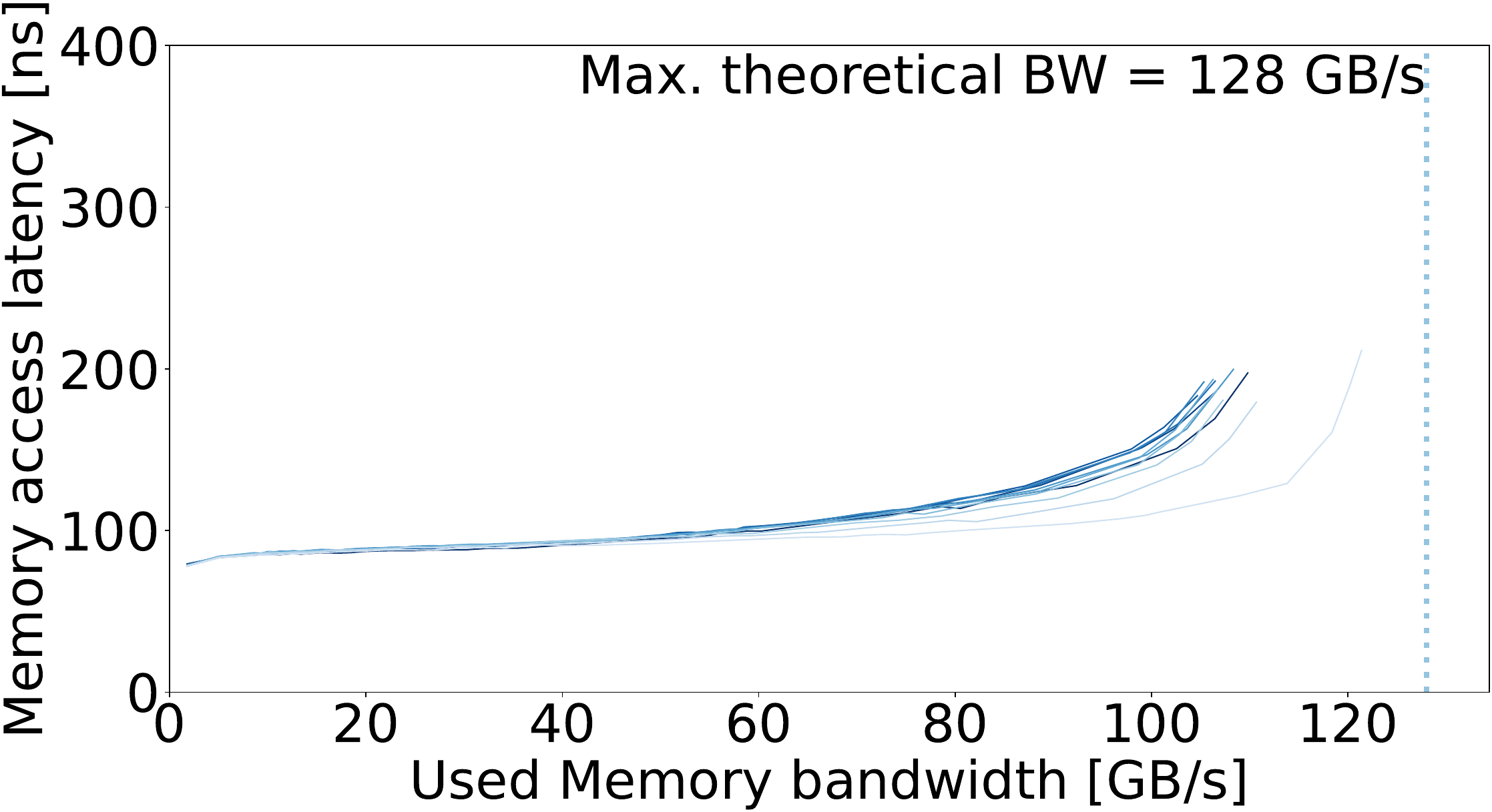}%
  }%

  \caption[Four subfigures]{%
    Broader applicability of our simulation-infrastructure enhancements.}
  \label{fig:final-mem-simulators}
\end{figure}

To demonstrate the broader  applicability of all our simulation enhancements, 
we deploy them in ZSim connected to Ramulator\,2 and DRAMsim3.  
The performance results are shown in Fig.\,\ref{fig:final-mem-simulators}.

After all corrections, the memory latency of all simulators under study
is still below the actual-system measurements.  
In the saturated memory system, the difference is large, up to 214\,ns. 
As a part of future work, it would be interesting to explore whether higher memory traffic intensity, 
e.g. caused by more aggressive data prefetchers, 
could further saturate the simulated memory systems leading to higher latencies.   
 
Another source of the latency error is the time memory requests spend in the memory controller. 
While the memory simulators under study implement the memory-controller functionalities, 
request selection, arbitration, scheduling decisions, etc., 
they do not model the time spent in the memory-controller pipeline, nor the latency of the memory PHY and IO~\cite{hansson:memctrl}. 
A simple way to correct these timings would be to include a delay-buffer that would increase the simulated  unloaded latency to match the actual one. 
More complex modeling would require detailed analysis of the memory controller designs  
and physical CPU-to-memory interconnects.

\section{Conclusion}
\label{sec:Conclusion}

This study addresses fundamental 
issues in memory-simulation accuracy in state-of-the-art simulation infrastructures. 
First, we propose a 
methodology that evaluates memory simulation performance from the memory simulator, CPU--memory interface and application perspectives. 
Second, we reveal significant discrepancies between internal simulator statistics and application-level performance, and find their root causes 
in CPU--memory interface, address mapping, NOC models and data prefetchers.   
Third,  we propose and implement fixes for these issues,
and reach close-to-hardware simulation accuracy. 
Evaluation across Ramulator, Ramulator\,2, and DRAMsim3 integrated with ZSim demonstrates that these changes are widely applicable. All corrections are released as open source.

\section*{Acknowledgments}

The authors would like to acknowledge John McCalpin from Barcelona Supercomputing Center for his valuable technical comments on NOC implementation of Intel Skylake CPUs.

\bibliographystyle{unsrt}
\bibliography{ref}

\newpage

\renewcommand{\appendixname}{Artifact Appendix}
\appendix

\section{{Artifact Appendix}}
\label{sec:Artifact}

\subsection{Abstract}

This artifact includes the source code and data required to replicate all experiments conducted in our study. We also provide the \texttt{00-damov-native} experiment to demonstrate that the inaccuracies identified in this paper also exist in the original DAMOV platform. This artifact enables readers to understand how the results were obtained, reproduce the experiments on their own systems, compare alternative configurations, and independently evaluate the behavior of the proposed interface beyond our specific platform. 
This artifact is organized as \texttt{Zsim-mem-Interface}.
It includes the simulator source trees used in the paper
(\texttt{zsim-bsc}, \texttt{DRAMsim3}, \texttt{Ramulator},
and \texttt{Ramulator2}), the benchmarks
(\texttt{ptr\_chase} and \texttt{traffic\_gen}),
experiment-specific configuration files, committed processed
outputs, and helper scripts to rerun or compare stages.

The repository is stage-oriented: each experiment fol   der
corresponds to one refinement step discussed in the paper
and maps directly to one or more figures. The artifact
covers the progression from the baseline interface
(Figure~2), through the clocking and model-correction stages
(Figures~3--5), to address mapping, NoC, and prefetcher
refinements (Figures~6a--6c), and finally the portability
experiments with DRAMsim3 and Ramulator2
(Figures~7c--7d). For each stage that produces paper
figures, the repository stores the processed
bandwidth--latency CSV and the corresponding plots.
Raw simulation traces are too large for Git; they are
hosted externally and referenced by per-experiment manifest
files.

\subsection{Artifact check-list (meta-information)}

{\small
\begin{itemize}
  \item {\bf Program:} ZSim-based CPU--memory simulation
    platform with Ramulator, Ramulator2, and DRAMsim3
    backends; pointer-chasing and traffic-generation
    benchmarks.
  \item {\bf Compilation:} GCC~11 or later (C++20
    required by Ramulator2), \texttt{scons},
    \texttt{make}, and Python\,3.
  \item {\bf Data set:} Committed processed CSV and PDF
    outputs for all figure-producing stages; externally
    hosted raw \texttt{bw-lat} trees referenced by
    per-experiment manifests.
  \item {\bf Run-time environment:} Linux x86\_64,
    Intel Pin 2.14 (build 71313), HDF5 1.8.16, Python
    with \texttt{pandas} and \texttt{matplotlib}.
  \item {\bf Hardware:} A Linux x86\_64 server or
    workstation. A multicore machine is recommended for
    rerunning full sweeps; inspection of committed outputs
    requires only local file access.
  \item {\bf Metrics:} Memory latency in nanoseconds;
    memory bandwidth in GB/s.
  \item {\bf Output:} Per-stage CSV tables and
    bandwidth--latency curves for the memory simulator
    view, memory interface view, and application view.
  \item {\bf Experiments:} Ten stage folders
    (\texttt{00-system-agnostic} through
    \texttt{09-portability-ramulator2}), corresponding to
    Figures~2--7.
  \item {\bf How much disk space required
    (approximately)?:} About 24\,GB for the full
    environment including simulator sources, benchmarks,
    and local raw-data mirrors.
  \item {\bf How much time is needed to prepare workflow
    (approximately)?:} 30--60 minutes once the external
    dependencies are available.
  \item {\bf How much time is needed to complete
    experiments (approximately)?:} Inspecting committed
    results takes minutes; regenerating plots from a raw
    \texttt{bw-lat} tree takes minutes to tens of minutes;
    full stage reruns require substantially more time.
  \item {\bf Publicly available:} Yes, via the artifact
    repository and externally hosted raw-data packages.
  \item {\bf Archived (provide DOI)?:} 10.5281/zenodo.19629351 
\end{itemize}
}

\subsection{Description}

The artifact is organized around the refinement sequence
presented in the paper. All experiment stages share the same
simulator sources and benchmarks; stages differ only in
their \texttt{sb.cfg} configuration and a small number of
stage-specific overrides. This design makes it possible to
compare intermediate states directly without duplicating the
codebase.

\subsubsection{How to access}

The repository is available at
\url{https://github.com/bsc-mem/ZSim-mem-Interface}.
The raw \texttt{bw-lat} packages are not committed to
Git; per-stage \texttt{raw-manifest.csv} files record
their download URLs and checksums.

\begin{figure}[!t]
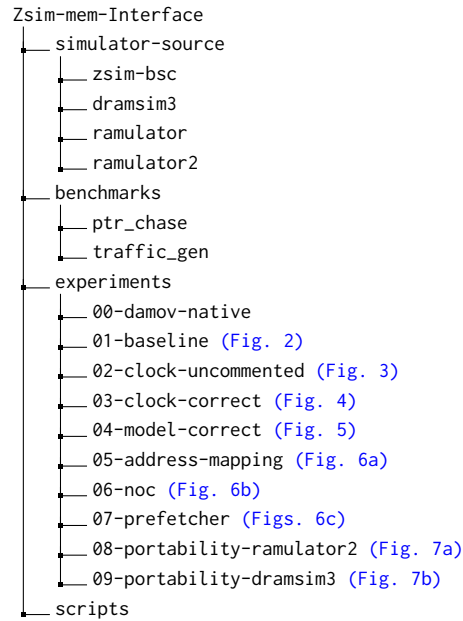

\footnotesize
\dirtree{%
.1 Zsim-mem-Interface.
.2 simulator-source.
.3 zsim-bsc.
.3 dramsim3.
.3 ramulator.
.3 ramulator2.
.2 benchmarks.
.3 ptr\_chase.
.3 traffic\_gen.
.2 experiments.
.3 00-damov-native.
.3 01-baseline \textcolor{blue}{(Fig.~2)}.
.3 02-clock-uncommented \textcolor{blue}{(Fig.~3)}.
.3 03-clock-correct \textcolor{blue}{(Fig.~4)}.
.3 04-model-correct \textcolor{blue}{(Fig.~5)}.
.3 05-address-mapping \textcolor{blue}{(Fig.~6a)}.
.3 06-noc \textcolor{blue}{(Fig.~6b)}.
.3 07-prefetcher \textcolor{blue}{(Figs.~6c)}.
.3 08-portability-ramulator2 \textcolor{blue}{(Fig.~7a)}.
.3 09-portability-dramsim3 \textcolor{blue}{(Fig.~7b)}.
.2 scripts.
}
\caption{Top-level structure of the artifact repository.}
\label{fig:artifact-tree}
\end{figure}

\subsubsection{Hardware dependencies}

The artifact targets a Linux x86\_64 environment. Intel
Pin~2.14 and the \texttt{ptr\_chase} benchmark are
Linux-specific, so the full workflow is not expected to run
on macOS or Windows. A single machine is sufficient for the
default artifact workflow; multicore hardware is beneficial
for full stage reruns.

\subsubsection{Software dependencies}

See the README inside \texttt{simulator-source/}
for all dependency details. The paper used
GCC~11.4.0; GCC~11 or later is required because
Ramulator2 needs C++20 support.
Other requirements are Intel Pin~2.14
(build~71313), HDF5~1.8.16, and Python~3 with
\texttt{pandas} and \texttt{matplotlib}. The build
system also requires the environment variables
\texttt{PINPATH}, \texttt{HDF5\_HOME},
\texttt{DRAMSIM3PATH}, \texttt{RAMULATORPATH}, and
\texttt{RAMULATOR2PATH}.

\subsubsection{Data sets}

For each stage that produces paper figures, the repository
commits:
\begin{itemize}
  \item the stage configuration (\texttt{sb.cfg} and
    optional overrides),
  \item the processed CSV
    (\texttt{bandwidth\_latency.csv}),
  \item the generated plots under \texttt{figures/}, and
  \item a \texttt{raw-manifest.csv} file that points to the
    externally hosted raw \texttt{bw-lat} tree.
\end{itemize}

The committed figures use the same terminology as the paper:
memory simulator view, memory interface view, and
application view. The portability stages provide the memory
interface and application views; the memory simulator view
is omitted when the corresponding data is unavailable.

\subsection{Installation}

After cloning the artifact repository, create a local
\texttt{.zsim-env} file with the required dependency paths,
source it, build ZSim, and build the benchmarks:

\begin{lstlisting}[language=sh, frame=single,
  basicstyle=\scriptsize, columns=fullflexible]
git clone https://github.com/bsc-mem/ZSim-mem-Interface.git
cd Zsim-mem-Interface

# edit .zsim-env to define PINPATH, HDF5_HOME,
# DRAMSIM3PATH, RAMULATORPATH, and RAMULATOR2PATH
source .zsim-env

cd simulator-source/zsim-bsc
scons --r -j$(nproc)
cd ../..

./scripts/build-benchmarks.sh
\end{lstlisting}

This preparation step is sufficient to inspect committed
results and to regenerate figures from an available raw
\texttt{bw-lat} tree.

\subsection{Experiment workflow}

The default artifact workflow is stage-based. Reviewers
can first inspect the committed outputs under each
stage's \texttt{processed/} and \texttt{figures/}
directories, then regenerate the same outputs from a raw
\texttt{bw-lat} tree, and finally compare stages.

For example, the baseline stage corresponding to Figure~2
can be exercised as follows:

\begin{lstlisting}[language=sh, frame=single,
  basicstyle=\scriptsize]
# summarize committed outputs for Figure 2
./scripts/reproduce-paper-results.sh 01-baseline

# regenerate processed outputs and plots
./experiments/plot.py ./raw-results/01-baseline/bw-lat \
  --config-dir ./experiments/01-baseline

# compare the baseline against the model-correct stage
./scripts/compare-results.sh 01-baseline 04-model-correct
\end{lstlisting}

The plotter writes regenerated files under
\texttt{test-output/} by default, so it does not
overwrite committed outputs. A full stage rerun is
started with \texttt{runner.sh} inside
\texttt{experiments/}; it expands the parameter sweep
and invokes \texttt{run-one.sh} for each point.

\subsection{Evaluation and expected results}

The main validation criterion is that regenerated
outputs match the committed ones. For the baseline
example, the CSV and plots under
\texttt{test-output/01-baseline/} should match those
under \texttt{experiments/01-baseline/}, and the
figures should reproduce the three views from
Figure~2.

Stages can also be compared directly.
\texttt{compare-results.sh} reports the deltas
between two stages (e.g., \texttt{01-baseline} and
\texttt{04-model-correct}), reflecting the interface
refinements discussed in the paper.

\subsection{Notes}

All interface refinements are controlled through ZSim
configuration files, except the Skylake-specific address
mapping used in Figure~6a, which is selected through a
dedicated Ramulator configuration file rather than
through manual source editing. This makes the
address-mapping stage fully reproducible without code
modifications.

\end{document}